\documentclass[11pt,showpacs,preprintnumbers,amsmath,amssymb,prd,nofootinbib,superscriptaddress]{revtex4-2}

\usepackage{dcolumn}
\usepackage{bm}
\usepackage{ifpdf}
\usepackage{hyperref}
\usepackage{bm}
\usepackage{xcolor,color,graphicx,graphics}
\usepackage[spanish,english]{babel}
\usepackage[latin1]{inputenc}
\usepackage[OT1]{fontenc}
\usepackage{latexsym,amssymb,amsmath,amsfonts}
\usepackage{makeidx}
\usepackage{epsfig,subfigure}
\usepackage{natbib}
\usepackage{epstopdf}
\usepackage{mathrsfs}
\usepackage{hyperref}
\hypersetup{colorlinks=true, linkcolor=blue, citecolor=blue, urlcolor=blue}
\usepackage{enumerate}

\everymath{\displaystyle}
\usepackage{graphicx}

\usepackage[T1]{fontenc}
\usepackage{amsmath}
\usepackage{amssymb}
\usepackage{graphicx}
\usepackage{xcolor}

\newcommand{\bea}{\begin{eqnarray}}
\newcommand{\eea}{\end{eqnarray}}

\newcommand{\orcid}[1]{\href{https://orcid.org/#1}{\includegraphics[width=10pt]{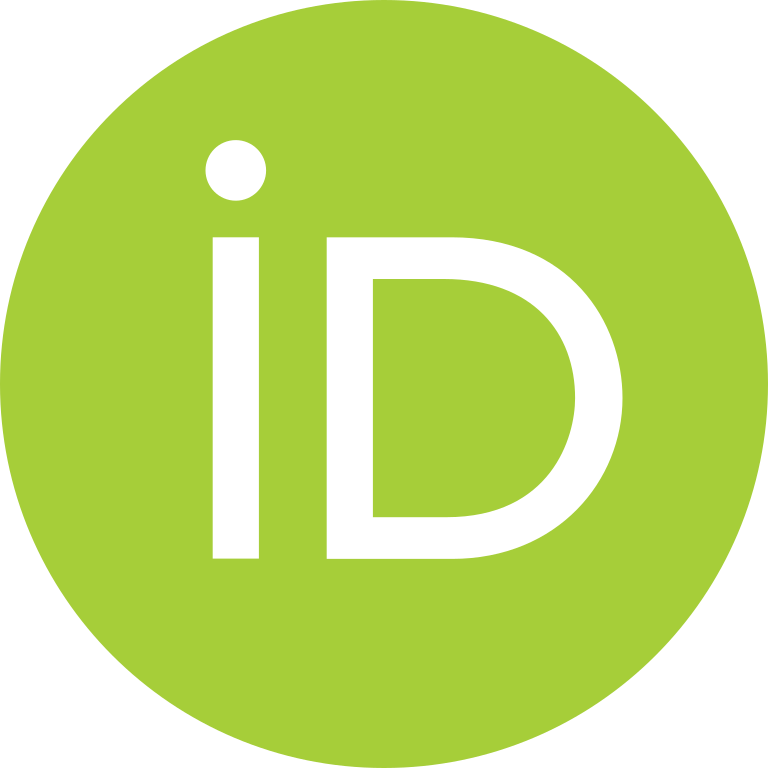}}}

\usepackage{fixmath}

\begin{document}

\title{Non-abelian aether-like term and applications at finite temperature}

\author{A. F. Santos \orcid{0000-0002-2505-5273}}
\email{alesandroferreira@fisica.ufmt.br}
\affiliation{Instituto de F\'{\i}sica, Universidade Federal de Mato Grosso,\\
78060-900, Cuiab\'{a}, Mato Grosso, Brazil}

\author{Faqir C. Khanna \orcid{0000-0003-3917-7578} \footnote{Professor Emeritus - Physics Department, Theoretical Physics Institute, University of Alberta\\
Edmonton, Alberta, Canada}}
\email{fkhanna@ualberta.ca; khannaf@uvic.ca}
\affiliation{Department of Physics and Astronomy, University of Victoria,\\
3800 Finnerty Road, Victoria BC V8P 5C2, Canada}

\begin{abstract}

The Yang-Mills-aether theory is considered. Implications of the  non-abelian aether-like term, which introduces violation of Lorentz symmetry, is investigated in a thermal quantum field theory. The Thermofield Dynamics formalism is used to introduce the temperature effects and spatial compactification. As a consequence, corrections due to the non-abelian aether term are calculated for the non-abelian Stefan-Boltzmann law and for the non-abelian Casimir energy and pressure at zero and finite temperature.


\end{abstract}

\maketitle

\section{Introduction} 

In recent years, attention has increased to the possibility that tiny violations of Lorentz symmetry may arise in theories that unify quantum mechanics and gravity \cite{KS, Pot}. This fundamental theory is expected to emerge at a very high energy scale. This symmetry breaking leads to the possibility that Lorentz symmetry might not be absolute, but only approximate. Lorentz symmetry violation opens up many opportunities to extend the well-known physics into new theories or corrections to standard physics. In order to explore the effects of this symmetry breaking, an effective theory has been constructed. The framework that describes all possible Lorentz violation is the Standard-Model Extension (SME) \cite{Kostelecky3, Kostelecky4, Kostelecky5}. The SME is composed of the general relativity and the standard model by adding all possible terms that violate Lorentz. Our interest is to study a non-Abelian extension of the Lorentz-breaking terms \cite{Kostelecky6}. There are some studies about these terms. For example, non-Abelian Carrol-Field-Jackiw term has been generated perturbatively \cite{JR, JR1}, some analyzes about the generations of this term have been investigated \cite{Sobreiro, Sobreiro1}, among others. Here the goal will be the non-abelian aether like term. The abelian aether term and its classical aspects has been analyzed \cite{Carroll} and its perturbative generation has been carried out \cite{Gomes}. The non-abelian version of the aether term in four-dimensions has been investigated and its perturbative generation has been performed \cite{Petrov, Petrov1}. There are studies that have not yet been development for the non-abelian aether term, such as the Casimir effect. Thus, additional theoretical predictions about the quantum vacuum are provided in this non-abelian extended theory. The study proposed here is an extension of the analysis developed in \cite{our}, where Lorentz invariant non-abelian theory is considered.

The Casimir effect can be interpreted as the force per unit area on bounding surfaces due to the confinement of a quantum field in a finite volume of space. As a consequence, two conducting parallel plates in the vacuum of a quantum field is attracted to each other due to boundary conditions, where the boundaries can be material media, interfaces between two phases of the vacuum, or topologies of space. This phenomenon was theoretically proposed by H. Casimir \cite{Casimir} and the first experimental observation was carried out by Sparnaay \cite{Sparnaay}. In this work the Casimir effect in the
vacuum of the non-abelian field with Lorentz violation is investigated. Since the Casimir force is a very sensitive measurement of quantum fluctuations of the field, it is a good way to investigated small Lorentz violating parameters. Furthermore, thermal corrections will be added to the Casimir effect for the non-abelian field in the presence of a non-abelian aether term. 

There are two ways to introduce temperature effects into quantum field theory: (i) imaginary-time formalism and (ii) real-time formalism. The first formalism was proposed by Matsubara \cite{Matsubara}, which is based on a substitution of time, $t$, by a complex time, $i\tau$. The second formalism is divided into two structure: the closed time path formalism \cite{Schwinger} and the Thermofield Dynamics (TFD) formalism \cite{Umezawa1, Umezawa2, Umezawa22, Khanna1, Khanna2}. The real-time formalism is applied to some studies where it is desirable to have an explicit dependence on time in addition to temperature. Here the TFD formalism is used. This approach depends on the doubling of the original Fock space and of the Bogoliubov transformation. The thermal space is composed of the original Fock space and a tilde (dual) space. The original and tilde space are related by a mapping, tilde conjugation rules, associating each operator ${\cal O}$ acting in Fock space with two operators in thermal Fock space. The Bogoliubov transformation is a rotation involving these two spaces, original and tilde, which introduce the temperature effects. TFD is a field theory on the topology $\Gamma_D^d=(\mathbb{S}^1)^d\times \mathbb{R}^{D-d}$ with $1\leq d \leq D$, where $D$ are the space-time dimensions and $d$ is the number of compactified dimensions. Here any set of dimensions of the manifold $\mathbb{R}^{D}$ can be compactified, where the circumference of the $nth$ $\mathbb{S}^1$ is specified by $\alpha_n$ that is the compactification parameter. Here three different topologies are considered. (1) The topology $\Gamma_4^1=\mathbb{S}^1\times\mathbb{R}^{3}$, where $\alpha=(\beta,0,0,0)$. In this case the time-axis is compactified in $\mathbb{S}^1$, with circumference $\beta$. In this topology the temperature effects are investigated. (2) The topology $\Gamma_4^1$ with $\alpha=(0,0,0,i2d)$, where the compactification along the coordinate $z$ is considered. Such a mechanism is quite suitable to treat, in particular, the Casimir effect. (3) The topology $\Gamma_4^2=\mathbb{S}^1\times\mathbb{S}^1\times\mathbb{R}^{2}$ with $\alpha=(\beta,0,0,i2d)$ is used. The double compactification allows the calculation of the Casimir effect at finite temperature.

This paper is organized as follows. In section II, the model that describes the non-abelian aether term is introduced. The non-abelian energy-momentum tensor and its vacuum expectation value is calculated. The 2, 3, and 4-point functions associated with the gluon field are analyzed. In section III, the TFD formalism is briefly presented. The energy-momentum tensor is written in terms of the compactification parameter. In section IV, the non-abelian Stefan-Boltzmann law and the non-abelian Casimir effect at zero and finite temperature are investigated. The effect due to the non-abelian aether term is discussed. In section V, some concluding remarks are presented.

\section{The Model - Non-abelian aether-like term}

The Lagrangian that describes the aether-like term that introduces a Lorentz-breaking Yang-Mills theory is
\bea
{\cal L}=\frac{1}{2}\mathrm{tr}\left(F_{\mu\nu}F^{\mu\nu}\right)+\lambda u^\mu u_\nu\mathrm{tr}\left(F_{\mu\lambda}F^{\nu\lambda}\right),\label{eq1}
\eea
where $F_{\mu\nu}=F_{\mu\nu}^aT^a$ is the non-abelian Lie algebra value field strength tensor, $\lambda$ is a parameter and $u_\mu$ is the constant Lorentz-violating vector which is dimensionless in four dimensions. Using that $\mathrm{tr}\left(T^aT^b\right)=-\frac{1}{2}\delta^{ab}$ the Lagrangian (\ref{eq1}) becomes
\bea
{\cal L}=-\frac{1}{4}F_{\mu\nu}^aF^{\mu\nu a}-\frac{\lambda}{2}u^\mu u_\nu F_{\mu\lambda}^a F^{\nu\lambda a}\label{eq2}
\eea
with $F^{\mu\nu a}= \partial^\mu A^{\nu a}-\partial^\nu A^{\mu a}-ef^{abc}A^{\mu b}A^{\nu c}$ being the non-Abelian field strength tensor.

In order to investigate applications of this non-abelian aether term at finite temperature, the energy-momentum tensor associated with such theory is calculated. It is defined as
\bea
T^{\mu\nu}=\frac{\partial{\cal L}}{\partial(\partial_\mu A_\lambda^a)}\partial^\nu A_\lambda^a-\eta^{\mu\nu}{\cal L}.
\eea
Using eq. (\ref{eq2}) the energy-momentum tensor becomes
\bea
T^{\mu\nu}=T^{\mu\nu}_{YM}+T^{\mu\nu}_{aether},
\eea
where $T^{\mu\nu}_{YM}$ is the Yang-Mills standard part, which is given by
\bea
T^{\mu\nu}_{YM}=-F^{\mu a}_\lambda F^{\nu\lambda a}+\frac{1}{4}\eta^{\mu\nu}F_{\rho\sigma}^aF^{\rho\sigma a}\label{SEMT}
\eea
and $T^{\mu\nu}_{aether}$ is new part corresponding to the non-abelian aether term given as
\bea
T^{\mu\nu}_{aether}=\frac{\lambda}{2}u^\alpha\left[\eta^{\mu\nu}u^\rho F_{\alpha\sigma}^aF_{\rho}^{\sigma a}-2\left(u^\mu F_{\alpha\sigma}^a-u_\sigma F_\alpha^{\mu a}\right)\partial^\nu A^{\sigma a}\right].\label{aether}
\eea
It should be noted that the standard part of the energy-momentum tensor given by eq. (\ref{SEMT}) is symmetric, but the aether part, eq. (\ref{aether}), is not symmetric. This is characteristic of theories where the Lorentz symmetry is violated.

To analyze problems and corrections due to the aether term, the vacuum expectation value of the energy-momentum tensor must be calculated. However, the vacuum state is not possible due to the product of field operators at the same point of the space-time. To avoid divergences, the energy-momentum tensor is written at different space-time points such as
\bea
T^{\mu\nu}(x)&=&\lim_{x'\rightarrow x}\tau\Bigl[-F^{\mu a}_\gamma(x)F^{\nu\gamma a}(x')+\frac{1}{4}\eta^{\mu\nu}F^a_{\rho\sigma}(x)F^{\rho\sigma a}(x')\nonumber\\
&+&\frac{\lambda}{2}u^\gamma\left(\eta^{\mu\nu}u^\rho F^a_{\gamma\sigma}(x)F^{\sigma a}_\rho(x')-2(u^\mu F_{\gamma\sigma}^a(x)-u_\sigma F_\gamma^{\mu a}(x))\partial^\nu A^{\sigma a}(x')\right)\Bigl],\label{EMT}
\eea
where $\tau$ is the time ordering operator. In order to use the quantization of the non-Abelian gauge fields, the canonical momenta is written as
\bea
\pi^{\mu a}=\frac{\partial{\cal L}}{\partial(\partial_0 A^a_\mu)}=-F^{0\mu,a},
\eea
and its components are
\bea
\pi^{0 a}=0\quad\quad\quad{\rm and}\quad\quad\quad\pi^{i a}=-F^{0i, a}.
\eea
Then the standard canonical commutation relation is given as
\bea
\left[A^a_\mu(x),\pi^b_\nu(x')\right]=i\delta^{ab}\,\delta_{\mu\nu}\,\delta^3(\vec{x}-\vec{x'}).
\eea
All other commutation relations are zero. It is important to note that the aether field is regarded as an interaction with the Yang-Mills field. Therefore, quantization is performed only in the pure Yang-Mills field.

Then the energy-momentum tensor becomes
\bea
T^{\mu\nu}(x)&=&\lim_{y,z,\omega\rightarrow x}\Bigl\{C_0^{\mu\nu}+C^{\mu\nu}+\left(-\Delta^{\mu\nu,\gamma\epsilon}+\lambda Z^{\mu\nu,\gamma\epsilon}\right)\tau\left[A^a_\gamma(x)A^a_\epsilon(y)\right]\nonumber\\
&+&gf^{abc}\left(\Delta^{\mu\nu}_{\gamma\sigma\Lambda}+\lambda K^{\mu\nu}_{\gamma\sigma\Lambda}\right)\tau\left[A^{\gamma a}(x)A^{\Lambda b}(y)A^{\sigma c}(z)\right]\nonumber\\
&+&\frac{1}{2}g^2f^{abc}f^{ade}\left(2\Delta^{\mu\nu}_{\gamma\Lambda\delta\rho}+\lambda\eta^{\mu\nu}\eta_{\Lambda\rho}u_\gamma u_\delta\right)\tau\left[A^{\gamma b}(x)A^{\Lambda c}(y)A^{\delta d}(z)A^{\rho e}(\omega)\right]\Bigl\},
\eea
where
\bea
\Delta^{\mu\nu,\gamma\epsilon}&\equiv& \Gamma^{\mu\delta,\nu}\,_\delta\,^{,\gamma\epsilon}-\frac{1}{4}\eta^{\mu\nu}\Gamma_{\delta\rho,}\,^{\delta\rho,\gamma\epsilon},\\
\Delta^{\mu\nu}_{\gamma\sigma\Lambda}&\equiv& -\Gamma^\mu_{\nu\gamma}\eta^\nu_\sigma+\frac{1}{4}\eta^{\mu\nu}\Gamma_{\sigma\Lambda\gamma}-\Gamma'^\nu_{\Lambda\gamma}\eta^\mu_\sigma+\frac{1}{4}\eta^{\mu\nu}\Gamma'_{\sigma\Lambda\gamma},\\
\Delta^{\mu\nu}_{\gamma\Lambda\delta\rho}&\equiv& -\eta^\mu_\gamma\eta^\nu_\delta\eta_{\Lambda\rho}+\frac{1}{4}\eta^{\mu\nu}\eta_{\Lambda\rho}\eta_{\gamma\delta},\\
Z^{\mu\nu,\gamma\epsilon}&\equiv& \frac{1}{2}\eta^{\mu\nu}u^\delta u^\rho\Gamma_{\delta\sigma,\rho}\,^{\sigma,\gamma\epsilon}-u^\delta u^\mu\Pi^{\nu\gamma\epsilon}\,_\delta+u^\delta u_\sigma\Pi^{\mu\gamma\sigma\epsilon}\,_\delta\,^\nu,\\
K^{\mu\nu}_{\gamma\sigma\Lambda}&\equiv& \frac{1}{2}\eta^{\mu\nu}u^\delta u^\rho\left(\Gamma_{\delta\sigma\gamma}\eta_{\rho\Lambda}-\Gamma'_{\rho\sigma\gamma}\eta_{\delta\Lambda}\right)+\left(u_\Lambda u^\mu\eta_{\gamma\sigma}-u_\Lambda u_\gamma \eta^\mu_\sigma\right)\partial'^\nu,\\
C_0^{\mu\nu}&\equiv& -i(n^\mu_0n^\nu_0-n_{0\delta}n^\delta_0)\delta^4(x-y),\\
C^{\mu\nu}&\equiv& \frac{1}{2}\eta^{\mu\nu}u^\delta u^\rho I_{\rho\alpha}-u^\delta u^\mu{\cal N}^\nu_\delta+u^\delta u_\sigma {\cal N}^{\nu\mu\sigma}\,_\delta,
\eea
with
\bea
\Gamma^{\mu\nu,\sigma\rho,\gamma\epsilon}&\equiv& (\eta^{\nu\gamma}\partial^\mu-\eta^{\mu\gamma}\partial^\nu)(\eta^{\rho\epsilon}\partial'^\sigma-\eta^{\sigma\epsilon}\partial'^\rho),\\
\Gamma^{\mu\nu\gamma}&\equiv& \eta^{\mu\gamma}\partial^\nu-\eta^{\nu\gamma}\partial^\mu,\\
\Pi^{\nu\gamma\epsilon\delta}&\equiv& \eta^{\gamma\epsilon}\partial^\delta\partial'^\nu-\eta^{\delta\gamma}\partial^\epsilon\partial'^\nu,\\
\Pi^{\mu\gamma\sigma\epsilon\delta\nu}&\equiv& \eta^{\mu\gamma}\eta^{\sigma\epsilon}\partial^\delta\partial'^\nu-\eta^{\gamma\delta}\eta^{\sigma\epsilon}\partial^\mu\partial'^\nu,\\
I^{\rho\delta}&\equiv& [A_\sigma^a(x),\partial'^\sigma A^{\rho a}(x')]n_0^\delta\delta({x_0-x'_0})-[A^{\delta^a}(x),\partial'^\sigma A^{\rho a}(x')]n_{0\sigma}\delta({x_0-x'_0}),\\
{\cal N}^\nu_\delta&\equiv& in^\nu_0(\delta^\sigma_\sigma n_{0\delta}-\delta^\sigma_\delta n_{0\sigma})\delta({\vec{x}-\vec{x'}}),\\
{\cal N}^{\nu\mu\sigma\delta}&\equiv& in^\nu_0(\delta^{\sigma\mu} n_0^\delta-\delta^{\sigma\delta} n_{0}^\mu)\delta({\vec{x}-\vec{x'}}),
\eea
and $n^\mu_0=(1,0,0,0)$ is a time-like vector.

Now, let us calculate the vacuum expectation value of the energy-momentum tensor which is defined as
\bea
\bigl\langle T^{\mu\nu}(x)\bigl\rangle&=& \bigl\langle 0| T^{\mu\nu}(x)|0\bigl\rangle,
\eea
where $|0\bigl\rangle$ is the ground state. Applying this definition to the energy-momentum tensor of the non-abelian field with contributions of the aether term, we obtain
\bea
\bigl\langle T^{\mu\nu}(x)\bigl\rangle&=&\lim_{y,z,\omega\rightarrow x}\Bigl\{\left(C_0^{\mu\nu}+C^{\mu\nu}\right)\bigl\langle 0|0\bigl\rangle+\left(-\Delta^{\mu\nu,\gamma\epsilon}+\lambda Z^{\mu\nu,\gamma\epsilon}\right)\Bigl\langle 0|\tau\left[A^a_\gamma(x)A^a_\epsilon(y)\right]|0\Bigl\rangle\nonumber\\
&+&gf^{abc}\left(\Delta^{\mu\nu}_{\gamma\sigma\Lambda}+\lambda K^{\mu\nu}_{\gamma\sigma\Lambda}\right)\Bigl\langle 0|\tau\left[A^{\gamma a}(x)A^{\Lambda b}(y)A^{\sigma c}(z)\right]|0\Bigl\rangle\nonumber\\
&+&\frac{1}{2}g^2f^{abc}f^{ade}\left(2\Delta^{\mu\nu}_{\gamma\Lambda\delta\rho}+\lambda\eta^{\mu\nu}\eta_{\Lambda\rho}u_\gamma u_\delta\right)\Bigl\langle 0|\tau\left[A^{\gamma b}(x)A^{\Lambda c}(y)A^{\delta d}(z)A^{\rho e}(\omega)\right]|0\Bigl\rangle\Bigl\}.\label{zero}
\eea
Here it is possible to define:
\begin{enumerate}
\item Two-points function:
\bea
\Bigl\langle 0\Bigl|\tau\left[A^a_\gamma(x)A^a_{\epsilon}(y)\right]\Bigl|0\Bigl\rangle&\equiv&\delta^{ab}\Bigl\langle 0\Bigl|\tau\left[A^a_\gamma(x)A^b_{\epsilon}(y)\right]\Bigl|0\Bigl\rangle\nonumber\\
&=&i\delta^{ab}D^{ab}_{\gamma\epsilon}(x-y)
\eea
with
\bea
D^{ab}_{\gamma\epsilon}(x-y)=\delta^{ab}\,\eta_{\gamma\epsilon}\,G_0(x-y)
\eea
being the gluon propagator and $G_0(x-y)$ the massless scalar field propagator which is given as
\bea
G_0(x-y)=-\frac{i}{(2\pi)^2}\frac{1}{(x-y)^2-i\epsilon}.\label{G0}
\eea

\item Three-gluon vertex:
\bea
\Bigl\langle 0\Bigl|\tau\left[A^{a\gamma}(x)A^{b\Lambda}(y)A^{c\sigma}(z)\right]\Bigl|0\Bigl\rangle\equiv-ig f^{abc}G_3^{\gamma\Lambda\sigma}(x, y,z),
\eea
where
\bea
G_3^{\gamma\Lambda\sigma}(x, y,z)&=&\eta^{\gamma\Lambda}\left(\partial_x^\sigma-\partial_y^\sigma\right)\delta(y-z)\delta(x-z)\nonumber\\
&+&\eta^{\Lambda\sigma}\left(\partial_y^\gamma-\partial_z^\gamma\right)\delta(z-x)\delta(y-x)\nonumber\\
&+&\eta^{\sigma\gamma}\left(\partial_z^\Lambda-\partial_x^\Lambda\right)\delta(x-y)\delta(z-y).
\eea

\item Four-gluon vertex:
\end{enumerate}
\bea
\Bigl\langle 0\Bigl|\tau\left[A^{b\gamma}(x)A^{c\Lambda}(y)A^{d\delta}(z)A^{e\rho}(\omega)\right]\Bigl|0\Bigl\rangle\equiv-ig^2\, G_4^{bcde, \gamma\Lambda\delta\rho}(x, y, z,\omega)
\eea
with
\bea
G_4^{bcde, \gamma\Lambda\delta\rho}(x, y, z,\omega)&=&\Bigl[f^{bcf}f^{def}\left(\eta^{\gamma\delta}\eta^{\rho\Lambda}-\eta^{\gamma\rho}\eta^{\Lambda\delta}\right)+f^{bdf}f^{cef}\left(\eta^{\gamma\rho}\eta^{\Lambda\delta}-\eta^{\gamma\Lambda}\eta^{\delta\rho}\right)\nonumber\\
&+&f^{bef}f^{cdf}\left(\eta^{\gamma\Lambda}\eta^{\delta\rho}-\eta^{\gamma\delta}\eta^{\rho\Lambda}\right)\Bigl]\delta(z-\omega)\delta(z-x)\delta(z-y).
\eea
Using these definitions, our objective is to obtain the energy-momentum tensor (\ref{zero}) in the presence of finite temperature. To achieve this goal, the TFD formalism will be introduced in the next section.

\section{TFD formalism}

Here a briefly introduction to the TFD formalism is presented. TFD is a quantum field theory at finite temperature where the statistical average of any operator ${\cal O}$ is interpreted as an expectation value in a thermal vacuum. This formalism is composed of two ingredients: (i) doubling of the original Hilbert space and (ii) the Bogoliubov transformation. The doubling is defined by the tilde ($^\thicksim$) conjugation rules which associates each operator in original Hilbert space, ${\cal S}$, to two operators in expanded space ${\cal S}_T$, where ${\cal S}_T={\cal S}\otimes \tilde{\cal S}$, with $\tilde{\cal S}$ being the tilde or dual space. 

The Bogoliubov transformation consists of a rotation in the tilde and non-tilde variables that introduces the effects of temperature. As an example, let us consider this transformation acting on a duplicated space, such as
\bea
\left( \begin{array}{cc} d(k,\alpha)  \\ \tilde d^\dagger(k,\alpha) \end{array} \right)={\cal B}(\alpha)\left( \begin{array}{cc} d(k)  \\ \tilde d^\dagger(k) \end{array} \right),
\eea
where ${\cal B}(\alpha)$ is the Bogoliubov transformation defined as
\bea
{\cal B}(\alpha)=\left( \begin{array}{cc} u(\alpha) & -v(\alpha) \\ 
-v(\alpha) & u(\alpha) \end{array} \right)
\eea
with $v(\alpha)=(e^{\alpha\omega}-1)^{-1/2}$ and $u(\alpha)=(1+v^2(\alpha))^{1/2}$ related to the Bose distribution. Here, the $\alpha$ parameter is assumed as the compactification parameter defined by $\alpha=(\alpha_0,\alpha_1,\cdots\alpha_{D-1})$, where $D$ are the space-time dimensions. For example, the effect of temperature is described by the choice $\alpha_0\equiv\beta$ and $\alpha_1,\cdots\alpha_{D-1}=0$.

In order to introduce the effects of temperature, let us write the n-point ($n=2,3,4$) Green function using the TFD formalism. The free scalar field propagator in a doublet notation is 
\bea
G_0^{(ab)}(x-y;\alpha)=i\langle 0,\tilde{0}| \tau[\phi^a(x;\alpha)\phi^b(y;\alpha)]| 0,\tilde{0}\rangle,
\eea
where $\phi(x;\alpha)={\cal B}(\alpha)\phi(x){\cal B}^{-1}(\alpha)$ and $a, b=1,2$. Then
\bea
G_0^{(ab)}(x-y;\alpha)=i\int \frac{d^4k}{(2\pi)^4}e^{-ik(x-y)}G_0^{(ab)}(k;\alpha),
\eea
where 
\bea
G_0^{(11)}(k;\alpha)\equiv G_0(k;\alpha)=G_0(k)+v^2(k;\alpha)[G_0(k)-G^*_0(k)],
\eea
with
\bea
G_0(k)=\frac{1}{k^2-m^2+i\epsilon}
\eea
and
\bea
[G_0(k)-G^*_0(k)]=2\pi i\delta(k^2-m^2).
\eea
The physical information is given by the component $a=b=1$, the non-tilde component. In addition, the generalized Bogoliubov transformation \cite{GBT} is given as
\bea
v^2(k;\alpha)=\sum_{s=1}^d\sum_{\lbrace\sigma_s\rbrace}2^{s-1}\sum_{l_{\sigma_1},...,l_{\sigma_s}=1}^\infty(-\eta)^{s+\sum_{r=1}^sl_{\sigma_r}}\,\exp\left[{-\sum_{j=1}^s\alpha_{\sigma_j} l_{\sigma_j} k^{\sigma_j}}\right],\label{BT}
\eea
with $d$ being the number of compactified dimensions, $\eta=1(-1)$ for fermions (bosons), $\lbrace\sigma_s\rbrace$ denotes the set of all combinations with $s$ elements and $k$ is the 4-momentum.

Similarly, the 3 and 4-point functions are given, respectively, as
\bea
G_3^{(11)\lambda\delta\Lambda}(k_1, k_2, k_3;\alpha)&=&G_3^{\lambda\delta\Lambda}(k_1, k_2, k_3)\nonumber\\
&+&v^2(k_1, k_2, k_3; \alpha)\left[G_3^{\lambda\delta\Lambda}(k_1, k_2, k_3)-G_3^{* \lambda\delta\Lambda}(k_1, k_2, k_3)\right]
\eea
and
\bea
&&G_4^{(11)bcde, \gamma\Lambda\delta\rho}(x, y, z,\omega;\alpha)=G_4^{bcde, \gamma\Lambda\delta\rho}(x, y, z,\omega)\nonumber\\
&+&v^2(k_1, k_2, k_3, k_4; \alpha)\left[G_4^{bcde, \gamma\Lambda\delta\rho}(k_1, k_2, k_3, k_4)-G_4^{* bcde, \gamma\Lambda\delta\rho}(k_1, k_2, k_3, k_4)\right].
\eea

Using this formalism, the $\alpha$ parameter is introduced in the vacuum expectation value of the energy-momentum tensor (\ref{zero}). Then
\bea
\bigl\langle T^{(ab)\mu\nu}(x;\alpha)\bigl\rangle&=&\lim_{y,z,\omega\rightarrow x}\Bigl\{\left(C_0^{(ab)\mu\nu}+C^{(ab)\mu\nu}\right)+i\,\Theta^{\mu\nu}G_0^{(ab)}(x-y;\alpha)\nonumber\\
&-&ig^2f^{abc}f^{abc}\left(\Delta^{\mu\nu}_{\gamma\sigma\Lambda}+\lambda K^{\mu\nu}_{\gamma\sigma\Lambda}\right)G_3^{(ab)\gamma\Lambda\sigma}(x, y,z;\alpha)\nonumber\\
&-& \frac{i}{2}g^4f^{abc}f^{ade}\left(2\Delta^{\mu\nu}_{\gamma\Lambda\delta\rho}+\lambda\eta^{\mu\nu}\eta_{\Lambda\rho}u_\gamma u_\delta\right)G_4^{(ab)bcde, \gamma\Lambda\delta\rho}(x, y, z,\omega;\alpha)\Bigl\},
\eea
where $\Theta^{\mu\nu}$ is defined as
\bea
\Theta^{\mu\nu}&=&-16\left(\partial_x^\mu\partial_y^\nu-\frac{1}{4}\eta^{\mu\nu}\partial_x^\rho\partial_{y\rho}\right)+8\lambda\Bigl[\frac{1}{2}\eta^{\mu\nu}u^\gamma u^\rho\left(2\partial_{x\gamma}\partial_{y\rho}+\eta_{\gamma\rho}\partial_{x\sigma}\partial_y^\sigma\right)\nonumber\\
&-&3u^\gamma u^\mu \partial_{x\gamma}\partial_y^\nu+u^\gamma u_{\sigma}\left(\eta^{\mu\sigma}\partial_{x\gamma}\partial_y^\nu-\eta_\gamma^\sigma\partial_x^\mu\partial_y^\nu\right)\Bigl].
\eea

To obtain a renormalized energy-momentum tensor, the Casimir prescription is carried out, i.e.
\bea
{\cal T}^{\mu\nu (ab)}(x;\alpha)=\langle T^{\mu\nu(ab)}(x;\alpha)\rangle-\langle T^{\mu\nu(ab)}(x)\rangle.
\eea
Thus
\bea
{\cal T}^{\mu\nu (ab)}(x;\alpha)&=&i\,\lim_{y,z,\omega\rightarrow x}\Bigl\{\Theta^{\mu\nu}\overline{G}_0^{(ab)}(x-y;\alpha)
-ig^2f^{abc}f^{abc}\left(\Delta^{\mu\nu}_{\gamma\sigma\Lambda}+\lambda K^{\mu\nu}_{\gamma\sigma\Lambda}\right)\overline{G}_3^{(ab)\gamma\Lambda\sigma}(x, y,z;\alpha)\nonumber\\
&-& \frac{i}{2}g^4f^{abc}f^{ade}\left(2\Delta^{\mu\nu}_{\gamma\Lambda\delta\rho}+\lambda\eta^{\mu\nu}\eta_{\Lambda\rho}u_\gamma u_\delta\right)\overline{G}_4^{(ab)bcde, \gamma\Lambda\delta\rho}(x, y, z,\omega;\alpha)\Bigl\}
\eea
with 
\bea
\overline{G}_0^{(ab)}(x-y;\alpha)&=&G_0^{(ab)}(x-y;\alpha)-G_0^{(ab)}(x-y),\\
\overline{G}_3^{(ab)\gamma\Lambda\sigma}(x, y,z;\alpha)&=&G_3^{(ab)\gamma\Lambda\sigma}(x, y,z;\alpha)-G_3^{(ab)\gamma\Lambda\sigma}(x, y,z),\\
\overline{G}_4^{(ab)bcde, \gamma\Lambda\delta\rho}(x, y, z,\omega;\alpha)&=&G_4^{(ab)bcde, \gamma\Lambda\delta\rho}(x, y, z,\omega;\alpha)-G_4^{(ab)bcde, \gamma\Lambda\delta\rho}(x, y, z,\omega).
\eea

Furthermore, the Fourier representation of the n-point functions are
\bea
\overline{G}_0^{(11)}(x-y;\alpha)=\int\frac{d^4k}{(2\pi)^4}e^{-ik(x-y)}v^2(k;\alpha)\left[G_0(k)-G_0^*(k)\right],
\eea
\bea
\overline{G}_3^{\lambda\delta\Lambda(11)}(x, y, z; \alpha)&=&\int\frac{d^4k_1}{(2\pi)^4}\frac{d^4k_2}{(2\pi)^4}\frac{d^4k_3}{(2\pi)^4}\,e^{-ik_1(y-z)}e^{-ik_2(y-x)}e^{-ik_3(x-z)}\nonumber\\
&\times & v^2(k_1, k_2, k_3; \alpha)\left[G_3^{\lambda\delta\Lambda}(k_1, k_2, k_3)-G_3^{* \lambda\delta\Lambda}(k_1, k_2, k_3)\right]
\eea
and a similar expression is obtained for the 4-point function.

Therefore, such development led the energy-momentum tensor to be written in the context of TFD formalism. Now, some applications are investigated for different choices of $\alpha$-parameter.

\section{Application at zero and finite temperature}

In this section three different topologies are considered. (i) The time-axis is compactified and $\alpha=(\beta,0,0,0)$.  (ii) The compactification along the coordinate $z$ is carried out and $\alpha=(0,0,0,i2d)$. (iii) The $\alpha$ parameter is choice as $\alpha=(\beta,0,0,i2d)$. In this case the double compactification consists in one being the time and the other along the coordinate $z$.

\subsection{Non-abelian thermal energy density}

For the first case, i.e. $\alpha=(\beta,0,0,0)$, the generalized Bogoliubov transformations are
\bea
v^2(k; \beta)&=&\sum_{j_0=1}^\infty e^{-\beta k j_0},\\
v^2(k_1, k_2, k_3; \beta)&=&\sum_{j_0=1}^\infty e^{-\beta (k_1+k_2+k_3) j_0},\\
v^2(k_1, k_2, k_3, k_4; \beta)&=&\sum_{j_0=1}^\infty e^{-\beta (k_1+k_2+k_3+k_4) j_0},
\eea
Taking these results to the Green functions, we get
\bea
\overline{G}_0^{(11)}(x-y;\beta)&=&\int \frac{d^4k}{(2\pi)^4}e^{-ik(x-y)}\sum_{j_0=1}^\infty e^{-\beta k^0 j_0}\left[G_0(k)-G_0^*(k)\right],\nonumber\\
&=&2\sum_{j_0=1}^\infty G_0\left(x-y-i\beta j_0 n_0\right),\label{1GF}
\eea
where $n_0^\mu=(1,0,0,0)$. Similar procedures are performed for the 3 and 4-point Green functions. Using these definitions, the energy-momentum tensor becomes
\bea
{\cal T}^{\mu\nu (11)}(x;\beta)&=&2i\,\lim_{y,z,\omega\rightarrow x}\sum_{j_0=1}^\infty\Bigl\{\Theta^{\mu\nu}G_0(x-y-i\beta j_0 n_0)\nonumber\\
&+&a\Bigl[({\cal D}_1^{\mu\nu}+\lambda {\cal E}_1^{\mu\nu})\delta(y-z-i\beta j_0 n_0)\delta(x-z-i\beta j_0 n_0)\nonumber\\
&+&({\cal D}_2^{\mu\nu}+\lambda {\cal E}_2^{\mu\nu})\delta(z-x-i\beta j_0 n_0)\delta(y-x-i\beta j_0 n_0)\nonumber\\
&+&({\cal D}_3^{\mu\nu}+\lambda {\cal E}_3^{\mu\nu})\delta(x-y-i\beta j_0 n_0)\delta(z-y-i\beta j_0 n_0)\Bigl]\nonumber\\
&+&\Phi^{\mu\nu}\delta(x-y-i\beta j_0 n_0)\delta(x-z-i\beta j_0 n_0)\delta(x-\omega-i\beta j_0 n_0)
\Bigl\},
\eea
with
\bea
a&=&-g^2f^{abc}f^{abc},\\
{\cal D}_1^{\mu\nu}&=&3\left[\partial_x^\mu(\partial_x^\nu-\partial_y^\nu)-\partial_y^\nu(\partial_x^\mu-\partial_y^\mu)-\frac{1}{4}\eta^{\mu\nu}(\partial_{x\sigma}-\partial_{y\sigma})(\partial_x^\sigma-\partial_y^\sigma)\right],\\
{\cal D}_2^{\mu\nu}&=&-(\partial_y^\mu-\partial_z^\mu)(\partial_x^\nu-\partial_y^\nu)+(\partial_x^\mu+\partial_y^\mu)(\partial_y^\nu-\partial_z^\nu),\\
{\cal D}_3^{\mu\nu}&=&-\eta^{\mu\nu}(\partial_{x\Lambda}-\partial_{y\Lambda})(\partial_z^\Lambda-\partial_x^\Lambda)+\partial_x^\mu(\partial_z^\nu-\partial_x^\nu)-\partial_y^\nu(\partial_z^\mu-\partial_x^\mu)\nonumber\\
&+&\frac{3}{4}\eta^{\mu\nu}(\partial_{x\Lambda}-\partial_{y\Lambda})(\partial_z^\Lambda-\partial_x^\Lambda),\\
{\cal E}_1^{\mu\nu}&=&\frac{1}{2}\eta^{\mu\nu}\left[u_\gamma u^\gamma(\partial_{x\sigma}+\partial_{y\sigma})(\partial_x^\sigma-\partial_y^\sigma)-u^\delta u_\sigma(\partial_{x\delta}+\partial_{y\delta})(\partial_x^\sigma-\partial_y^\sigma)\right]\nonumber\\
&+&u_\sigma u^\mu\partial_y^\nu(\partial_x^\sigma-\partial_y^\sigma)-u_\gamma u^\gamma\partial_y^\nu(\partial_x^\mu-\partial_y^\mu),\\
{\cal E}_2^{\mu\nu}&=&\frac{1}{2}\eta^{\mu\nu}\left[u_\gamma u^\sigma(\partial_{x\sigma}+\partial_{y\sigma})(\partial_y^\gamma-\partial_z^\gamma)-u^\delta u_\gamma(\partial_{x\delta}+\partial_{y\delta})(\partial_y^\gamma-\partial_z^\gamma)\right]\nonumber\\
&+&u_\gamma u^\mu\partial_y^\nu(\partial_x^\gamma-\partial_y^\gamma)-u^\mu u_\gamma\partial_y^\nu(\partial_y^\gamma-\partial_z^\gamma),\\
{\cal E}_3^{\mu\nu}&=&\frac{1}{2}\eta^{\mu\nu}\left[u^\sigma u_\Lambda(\partial_{x\sigma}+\partial_{y\sigma})(\partial_z^\Lambda-\partial_x^\Lambda)-4u^\delta u_\gamma(\partial_{x\delta}+\partial_{y\delta})(\partial_z^\gamma-\partial_x^\gamma)\right]\nonumber\\
&+&4u_\gamma u^\mu\partial_y^\nu(\partial_z^\gamma-\partial_x^\gamma)-u^\mu u_\gamma\partial_y^\nu(\partial_z^\gamma-\partial_x^\gamma),\\
\Phi^{\mu\nu}&=&-\frac{3}{2}\lambda g^4 f^{abc}f^{ade}\eta^{\mu\nu} u^\rho u_\rho\left(f^{bch}f^{deh}-f^{beh}f^{cdh}\right).
\eea

To investigate the corrections due to the non-abelian aether-like term, let us consider the vector $u^\mu$ as a time-like vector, i.e. $u^\mu=(1,0,0,0)$. Using this choice and $\mu=\nu=0$, after several calculations, the non-abelian energy density is given as
\bea
{\cal T}^{00}(T)&=&\frac{8\pi^2}{15}(1+\lambda)T^4+a(\frac{1}{2}-3\lambda)T^2\sum_{j_0=1}^\infty \frac{i}{j_0^2}\delta^2(-i\beta j_0)+b\lambda\sum_{j_0=1}^\infty i\delta^3(-i\beta j_0),\label{SBL}
\eea
where $b\equiv 3g^4 f^{abc}f^{ade}\left(f^{bch}f^{deh}-f^{beh}f^{cdh}\right)$. Here, it should be noted that the product of Dirac delta functions with identical arguments is not well defined. However, to avoid this problem the regularized form of delta function is defined as \cite{Van}
\bea
2\pi i\,\delta^n(x)=\left(-\frac{1}{x+i\epsilon}\right)^{n+1}-\left(-\frac{1}{x-i\epsilon}\right)^{n+1}.
\eea

The eq. (\ref{SBL}) is the non-abelian Stefan-Boltzmann law with corrections due to the non-abelian aether term which leads to Lorentz symmetry breaking. This correction affects all self-interactions of gluons. The temperature dependence is different for each contribution. In the high-temperature limit, the correction with dependency $T^4$  is dominant, while for low-temperature, the four-gluon self-interaction term is dominant. Furthermore, in the case where the non-abelian aether term is zero, the standard non-abelian Stefan-Boltzmann law for gluons is recovered. In order to visualize the corrections due to the aether term, in Figure 1 the energy as a function of temperature is plotted. However, as a first approximation, only the first term is considered, since the product of the Dirac delta functions is not well defined.
\begin{figure}[h]
\includegraphics[scale=0.8]{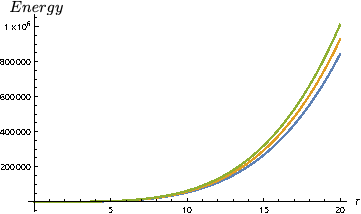}
\caption{Stefan-Boltzmann law with corrections due to the aether term. The blue curve presents the standard result, that is, the Lorentz invariant Stefan-Boltzmann law, while the yellow and green curves show the Stefan-Boltzmann law with corrections due to Lorentz violation for $\lambda=0.1$ and $\lambda=0.2$, respectively. }
\end{figure}

\subsection{Non-abelian Casimir effect with the non-abelian aether term}

Here, the corrections of the non-abelian aether term will be investigated for two cases: the Casimir effect at zero and finite temperature.

\subsubsection{Non-abelian Casimir effect at zero temperature}

In this case the $\alpha$-parameter is choice as $\alpha=(0,0,0,2id)$. This leads to the Bogoliubov transformations
\bea
v^2(k;d)&=&\sum_{l_3=1}^\infty e^{-i2dk l_3},\\
v^2(k_1, k_2, k_3; d)&=&\sum_{l_3=1}^\infty e^{-i2d (k_1+k_2+k_3) l_3},\\
v^2(k_1, k_2, k_3, k_4; d)&=&\sum_{l_3=1}^\infty e^{-i2d (k_1+k_2+k_3+k_4) l_3}.
\eea
Using these transformations in the Green functions, the energy-momentum tensor becomes
\bea
{\cal T}^{\mu\nu (11)}(x;d)&=&2i\,\lim_{y,z,\omega\rightarrow x}\sum_{j_0=1}^\infty\Bigl\{\Theta^{\mu\nu}G_0(x-y-2dl_3z)\nonumber\\
&+&a\Bigl[({\cal D}_1^{\mu\nu}+\lambda {\cal E}_1^{\mu\nu})\delta(y-z-2dl_3z)\delta(x-z-2dl_3z)\nonumber\\
&+&({\cal D}_2^{\mu\nu}+\lambda {\cal E}_2^{\mu\nu})\delta(z-x-2dl_3z)\delta(y-x-2dl_3z)\nonumber\\
&+&({\cal D}_3^{\mu\nu}+\lambda {\cal E}_3^{\mu\nu})\delta(x-y-2dl_3z)\delta(z-y-2dl_3z)\Bigl]\nonumber\\
&+&\Phi^{\mu\nu}\delta(x-y-2dl_3z)\delta(x-z-2dl_3z)\delta(x-\omega-2dl_3z)
\Bigl\}.
\eea
Taking $\mu=\nu=0$, the non-abelian Casimir energy at zero temperature is
\bea
{\cal T}^{00}(d)&=&-\frac{\pi^2}{90d^4}(1+\lambda)+\frac{a}{d^2}\left(-\frac{1}{8}+\frac{3}{4}\lambda\right)\sum_{l_3=1}^\infty \frac{i}{l_3^2}\delta^2(-2d l_3)+b\lambda\sum_{l_3=1}^\infty i\delta^3(-2d l_3).
\eea
And for $\mu=\nu=3$, the non-abelian Casimir pressure at zero temperature is
\bea
{\cal T}^{33}(d)&=&-\frac{\pi^2}{15d^4}\left(\frac{1}{2}+\frac{\lambda}{3}\right)+\frac{a}{d^2}\left(\frac{7}{4}-\frac{1}{4}\lambda\right)\sum_{l_3=1}^\infty \frac{i}{l_3^2}\delta^2(-2d l_3)-b\lambda\sum_{l_3=1}^\infty i\delta^3(-2d l_3).
\eea
Since the distance between the plates $d$ is small, the first term is dominant compared to the other terms that come from the self-interaction of the gluons. Furthermore, corrections due to the non-abelian aether term contribute to increase the Casimir force, which is an attractive force. Figure 2 shows the Casimir pressure, which is dependent on the aether term, as a function of the distance between the plates. Here only the first correction is considered due to the difficulty of numerically expressing the terms that relate to the Dirac delta functions and their sum.
\begin{figure}[h]
\includegraphics[scale=0.8]{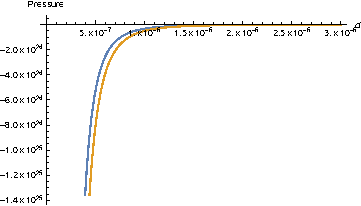}
\caption{Casimir pressure with corrections due the the aether term. The blue curve expresses the usual Casimir effect ($\lambda=0$) and the yellow curve shows the corrections to the Casimir pressure for the Lorentz violation case, considering $\lambda=0.8$ (this is a theoretical choice to better visualize the modifications caused by the aether term).	 }
\end{figure}

\subsubsection{Non-abelian Casimir effect at finite temperature}

To investigate both effects, finite temperature and spatial compactification, the $\alpha$-paramenter is given as $\alpha=(\beta, 0, 0, i2d)$. As a consequence, Bogoliubov transformations become
\bea
v^2(k;\beta,d)&=&v^2(k;\beta)+v^2(k;d)+2v^2(k;\beta)v^2(k;d),\nonumber\\
&=&\sum_{j_0=1}^\infty e^{-\beta kj_0}+\sum_{l_3=1}^\infty e^{-i2dkl_3}+2\sum_{j_0,l_3=1}^\infty e^{-\beta kj_0-i2dkl_3}\\
v^2(k_1, k_2, k_3;\beta,d)&=&\sum_{j_0=1}^\infty e^{-\beta (k_1+k_2+k_3) j_0}+\sum_{l_3=1}^\infty e^{-i2d (k_1+k_2+k_3) l_3}\nonumber\\
&+&2\sum_{j_0,l_3=1}^\infty e^{-\beta (k_1+k_2+k_3) j_0-i2d (k_1+k_2+k_3) l_3},\\
v^2(k_1, k_2, k_3, k_4;\beta,d)&=&\sum_{j_0=1}^\infty e^{-\beta (k_1+k_2+k_3+k_4) j_0}+\sum_{l_3=1}^\infty e^{-i2d (k_1+k_2+k_3+k_4) l_3}\nonumber\\
&+&2\sum_{j_0,l_3=1}^\infty e^{-\beta (k_1+k_2+k_3+k_4) j_0-i2d (k_1+k_2+k_3+k_4) l_3}.
\eea
In these expressions, the first term leads to the non-abelian Stefan Boltzmann law and the second term to the non-abelian Casimir effect at zero temperature. The combined effect of temperature and spatial compactification that leads to the non-abelian Casimir effect at finite temperature is given by the third term.

Then the energy-momentum tensor is given as
\bea
{\cal T}^{\mu\nu (11)}(x;\beta, d)&=&4i\,\lim_{y,z,\omega\rightarrow x}\sum_{j_0=1}^\infty\Bigl\{\Theta^{\mu\nu}G_0(x-y-i\beta j_0 n_0-2dl_3z)\nonumber\\
&+&a\Bigl[({\cal D}_1^{\mu\nu}+\lambda {\cal E}_1^{\mu\nu})\delta(y-z-i\beta j_0 n_0-2dl_3z)\delta(x-z-i\beta j_0 n_0-2dl_3z)\nonumber\\
&+&({\cal D}_2^{\mu\nu}+\lambda {\cal E}_2^{\mu\nu})\delta(z-x-i\beta j_0 n_0-2dl_3z)\delta(y-x-i\beta j_0 n_0-2dl_3z)\nonumber\\
&+&({\cal D}_3^{\mu\nu}+\lambda {\cal E}_3^{\mu\nu})\delta(x-y-i\beta j_0 n_0-2dl_3z)\delta(z-y-i\beta j_0 n_0-2dl_3z)\Bigl]\nonumber\\
&+&\Phi^{\mu\nu}\delta(x-y-i\beta j_0 n_0-2dl_3z)\delta(x-z-i\beta j_0 n_0-2dl_3z)\nonumber\\
&\times&\delta(x-\omega-i\beta j_0 n_0-2dl_3z)
\Bigl\}.
\eea
For $\mu=\nu=0$, we get
\bea
{\cal T}^{00(11)}(\beta;d)&=&\frac{32}{\pi^2}\sum_{j_0,l_3=1}^\infty\frac{3(\beta j_0)^2-(2dl_3)^2}{[(\beta j_0)^2+(2dl_3)^2]^3}(1+\lambda)\nonumber\\
&+&2a\left(-\frac{1}{2}+3\lambda\right)\sum_{j_0,l_3=1}^\infty\frac{i\delta^2(-i\beta j_0-2dl_3)}{(i\beta j_0+2dl_3)^2}+2b\lambda\sum_{j_0,l_3=1}^\infty i\delta^3(-i\beta j_0-2dl_3).\label{ED}
\eea
This is the non-abelian Casimir energy at finite temperature. Choosing $\mu=\nu=3$ leads to
\bea
{\cal T}^{33(11)}(\beta;d)&=&\frac{32}{\pi^2}\sum_{j_0,l_3=1}^\infty\frac{(\beta j_0)^2-3(2dl_3)^2}{[(\beta j_0)^2+(2dl_3)^2]^3}(1+2\lambda)\nonumber\\
&+&2a(7-\lambda)\sum_{j_0,l_3=1}^\infty\frac{i\delta^2(-i\beta j_0-2dl_3)}{(i\beta j_0+2dl_3)^2}-2b\lambda\sum_{j_0,l_3=1}^\infty i\delta^3(-i\beta j_0-2dl_3).\label{PD}
\eea
This is the non-abelian Casimir pressure at finite temperature. It is interesting to note that the non-abelian aether term modifies the Casimir effect in both cases, at zero and finite temperature. Figure 3 shows the Casimir pressure at finite temperature with corrections due to the aether term. One can see the contributions of the aether term as a function of temperature. It is important to note that Figures 1, 2 and 3 present only a simple attempt to visualize the standard effect and how the Lorentz violation effects change these quantities. Furthermore, it is worth emphasizing that for a complete representation, more experimental results on this theory at finite temperature must be obtained, as well as more information about the numerical values of the product involving Dirac delta functions with the same arguments.
\begin{figure}[h]
\includegraphics[scale=0.8]{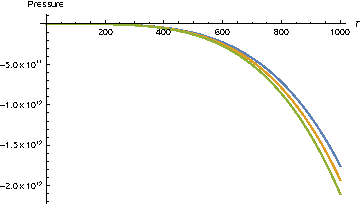}
\caption{Casimir pressure as a function of temperature with corrections due to the aether term. Here the distance between the plates $d$ is taken as $d=10^{-6}m$. The blue curve displays the Lorentz invariant phenomenon with $\lambda=0$, the yellow and green curves show the corrections due to the aether term for the cases $\lambda=0.1$ and $\lambda=0.2$, respectively.}
\end{figure}

\section{Conclusion}

High energy physics which discusses a fundamental theory with general relativity and standard model together leads to the possibility that tiny violations of Lorentz symmetry may arise. In this context, the Yang-Mills theory with a non-abelian aether term is considered. The energy-momentum tensor for the Yang-Mills theory with Lorentz violation is determined. Then some applications at finite temperature are investigated. The temperature effects are introduced using the TFD formalism, a real-time formalism for the thermal quantum field theory. First the expression for the energy-momentum tensor in terms of the TFD propagator considering a topology $\Gamma_4^1=\mathbb{S}^1\times\mathbb{R}^{3}$ is derived. This is used to analyze the non-abelian field with Lorentz violation at finite temperature, where the time coordinate is compactified. The Stefan-Boltzmann law with corrections due to the non-abelian aether term is obtained and the possible consequences due to this Lorentz-violating term are discussed. The next analysis is done with the compactification along of $z$ coordinates. Then the non-abelian Casimir energy and pressure are calculated. The non-abelian aether term contribute to increase the attractive Casimir force. The last investigation considered a topology $\Gamma_4^2=\mathbb{S}^1\times\mathbb{S}^1\times\mathbb{R}^{2}$. This was used to study the non-abelian field compactified in spatial coordinate and at finite temperature. In this case the Casimir energy and pressure at finite temperature are determined.  Therefore, the deviations of the Casimir force in the Lorentz violating extension of the Yang-Mills theory at zero and finite temperature have been calculated.

\section*{Acknowledgments}

This work by A. F. S. is supported by National Council for Scientific and Technological Develo\-pment - CNPq project No. 313400/2020-2.

\section*{Availability Statment}
This is a theoretical work and all previous results are listed in the references.


\begin{thebibliography}{100}

\bibitem{KS} V. A. Kostelecky and S. Samuel, Phys. Rev. D {\bf 39}, 683 (1989).
\bibitem{Pot} V. A. Kostelecky and R. Potting, Nucl. Phys. B {\bf 359}, 545 (1991).
\bibitem{Kostelecky3} D. Colladay and V. A. Kostelecky, Phys. Rev. D {\bf 55}, 6760 (1997).
\bibitem{Kostelecky4} D. Colladay and V. A. Kostelecky, Phys. Rev. D {\bf 58}, 116002 (1998).
\bibitem{Kostelecky5} V. A. Kostelecky, Phys. Rev. D {\bf 69}, 105009 (2004).
\bibitem{Kostelecky6} V. A. Kostelecky and Z. Li, Phys. Rev. D {\bf 99}, 056016 (2019).
\bibitem{JR} M. Gomes, J. R. Nascimento, E. Passos, A. Yu. Petrov and A. J. da Silva, Phys. Rev. D {\bf 76}, 047701 (2007).
\bibitem{JR1} T. Mariz, M. Gomes, J. R. Nascimento, A. Yu. Petrov and A. J. da Silva, Phys. Lett. B {\bf 661}, 312 (2008).
\bibitem{Sobreiro} T. R. S. Santos and R. F. Sobreiro, Phys. Rev. D {\bf 91}, 025008 (2015).
\bibitem{Sobreiro1} T. R. S. Santos and R. F. Sobreiro, Eur. Phys. J. C {\bf 77}, 903 (2017).
\bibitem{Carroll} S. Carroll, H. Tam, Phys. Rev. D {\bf 78}, 044047 (2008).
\bibitem{Gomes} M. Gomes, J. R. Nascimento, A. Y. Petrov and A. J. da Silva, Phys. Rev. D {\bf 81}, 045018 (2010).
\bibitem{Petrov} D. R. Granado, C. P. Felix, I. F. Justo, A. Yu. Petrov and D. Vercauteren, Annals of Physics {\bf 422}, 168324 (2020).
\bibitem{Petrov1} A. J. G. Carvalho, D. R. Granado, J. R. Nascimento and A. Yu. Petrov, Eur. Phys. J. C {\bf 79}, 817 (2019).
\bibitem{our} A. F. Santos and Faqir C. Khanna, Int. J. Mod. Phys. A {\bf 34}, 1950128 (2019).
\bibitem{Casimir} H. G. B. Casimir, Proc. K. Ned. Akad. Wet. {\bf 51}, 793 (1948).
\bibitem{Sparnaay} M. J. Sparnaay, Physica {\bf 24}, 751 (1958).
\bibitem{Matsubara} T. Matsubara, Prog. Theor. Phys. {\bf 14}, 351 (1955).
\bibitem{Schwinger}J. Schwinger, J. Math. Phys. {\bf 2}, 407 (1961); J. Schwinger, Lecture Notes Of Brandeis University
Summer Institute (1960).
\bibitem{Umezawa1}Y. Takahashi and H. Umezawa, Collective Phenomena 2, 55 (1975) (Reprinted in Int. J. Mod. Phys. B 10, 1755 (1996)).
\bibitem{Umezawa2}Y. Takahashi, H. Umezawa and H. Matsumoto, Thermofield Dynamics and Condensed States, North-Holland, Amsterdan, (1982); F. C. Khanna, A. P. C. Malbouisson, J. M. C. Malboiusson and A. E. Santana, Themal quantum field theory: Algebraic aspects and applications, World Scientific, Singapore, (2009).
\bibitem{Umezawa22} H. Umezawa, Advanced Field Theory: Micro, Macro and Thermal Physics, AIP, New York, (1993).
\bibitem{Khanna1} A. E. Santana and F. C. Khanna, Phys. Lett. A {\bf 203}, 68 (1995).
\bibitem{Khanna2} A. E. Santana, F. C. Khanna, H. Chu, and C. Chang, Ann. Phys. {\bf 249}, 481 (1996).
\bibitem{GBT}F. C. Khanna, A. P. C Malbouisson, J. M. C. Malbouisson and A. E. Santana, Ann. Phys. {\bf 326}, 2634 (2011).
\bibitem{Van} Ch.G. van Weert, {\it An Introduction to Real- and Imaginary-time Thermal Field Theory}, Lecture notes on Statistical Field Theory (2001).

\end{thebibliography}
\end{document}